\documentclass[aps]{revtex4}

\input{tcilatex}

\begin{document}

\title{NMR-like control of a quantum bit superconducting circuit}
\author{E. Collin, G.\ Ithier, A. Aassime, P. Joyez, D.\ Vion, and D.\ Esteve%
}
\affiliation{Quantronics group, Service de Physique de l'Etat Condens\'{e}, DSM/DRECAM,
CEA Saclay, 91191\ Gif-sur-Yvette, France}
\date{submitted 17-3}
\pacs{PACS number}

\begin{abstract}
Different nanofabricated superconducting circuits based on Josephson
junctions have already achieved a degree of quantum coherence sufficient to
demonstrate coherent superpositions of their quantum states. These circuits
are considered for implementing quantum bits, which are the building blocks
for the recently proposed quantum computer designs. We report experiments in
which the state of a quantum bit circuit, the quantronium, is efficiently
manipulated using methods inspired from Nuclear Magnetic Resonance (NMR).
\end{abstract}

\maketitle

Despite progress in the development of quantum bit (qubit) electronic
circuits, the complexity and robustness of the operations that have been
performed on them are still too primitive for implementing a small quantum
processor circuit that would demonstrate the power of quantum computing~
(see~\cite{NC} for a review).\ Presently, the most advanced qubit circuits
are superconducting ones based on Josephson junctions.\ The preparation of
coherent superpositions of the two states of a qubit has already been
achieved for several circuits~\cite%
{Nak2,VionSCIENCE,ACthesis,Martinis,delft,Chalmers,Buisson}, and a two qubit
gate was recently demonstrated~\cite{NECgate}. The coherence time of the
superpositions prepared is however short, which explains why qubit
transformations are far less developed for qubit circuits than for
microscopic\ quantum systems such as atoms or spins.\ In this Letter, we
report on experiments that successfully manipulate a Josephson qubit based
on the quantronium circuit~\cite{quantroniumPHYSC}, using methods developed
in NMR. We demonstrate that any transformation of the qubit can be
implemented, that they can be made robust against detuning effects and that
decoherence due to noise in the control parameters can be fought.

The quantronium circuit, described in Fig.~1, is derived from the Cooper
pair box~\cite{Bouchiat98, NakBOX}. It consists of a superconducting loop
interrupted by two adjacent small Josephson tunnel junctions with
capacitance $C_{j}/2$ and Josephson energy $E_{J}/2$ each, and by a larger
Josephson junction $(E_{J0}\approx 15E_{J})$ for readout. The two small
junctions define a low capacitance $C_{\Sigma }$ superconducting electrode
called the ``island'', with charging energy $E_{C}=(2e)^{2}/2C_{\Sigma }$.
This island is biased by a voltage source $U$ through a gate capacitance $%
C_{g}$. Fabrication is performed using standard e-beam lithography, and the
characteristic energies measured in the sample reported in this work are $%
E_{J}=0.87~k_{B}~\mathrm{K}$ and $E_{C}=0.66~k_{B}~\mathrm{K}$\textrm{.}
Experiments are performed at $20~\mathrm{mK}$ using filtered connecting
lines. The eigenstates of this system are determined by the dimensionless
gate charge $N_{g}=C_{g}U/2e$, and by the superconducting phase $\delta
=\gamma +\phi $ across the two small junctions, where $\gamma $ is the phase
across the large junction and $\phi =\Phi /\varphi _{0}$, with $\Phi $ the
external flux imposed through the loop and $\varphi _{0}=\hbar /2e$. For a
large range of parameters, the spectrum is anharmonic, and the two lowest
energy states $\left| 0\right\rangle $ and $\left| 1\right\rangle $ form a
two-level system suitable for a qubit.\ At the optimal working point ($%
\delta =0,N_{g}=1/2)$,\ the transition frequency $\nu _{01}$ is stationary
with respect to changes in the control parameters, which makes the
quantronium immune to phase and charge noise~\cite{quantroniumPHYSC,
VionSCIENCE}. For the sample investigated, $\nu _{01}\simeq 16.40~\mathrm{GHz%
}$\ at the optimal working point. For readout, $\left| 0\right\rangle $ and $%
\left| 1\right\rangle $ are discriminated through the difference in their
supercurrents in the loop~\cite{VionSCIENCE}. A trapezoidal readout pulse $%
I_{b}(t)$ with a peak value slightly below the critical current $%
I_{0}=E_{J0}/\varphi _{0}\simeq 550~\mathrm{nA}$ is applied to the circuit
so that,\ by adjusting the amplitude and duration of the \ pulse, the
switching of the large junction to a finite\emph{\ }voltage state is induced
with a large probability $p_{1}$ for state $\left| 1\right\rangle $ and with
a small probability $p_{0}$ for state $\left| 0\right\rangle $. The
switching is detected by measuring the voltage across the readout junction
with a room temperature amplifier, and the switching probability $p$ is
determined by repeating the experiment a few 10$^{4}$ times at a rate $10-60~%
\mathrm{kHz}$.\ The fidelity of the measurement is the largest value of $%
\eta =p_{1}-p_{0}$.

\ The manipulation of the qubit state is achieved by applying time dependent
control parameters $N_{g}(t)$ and $I_{b}(t)$. When a nearly resonant
microwave voltage is applied to the gate, the Hamiltonian is conveniently
described using the Bloch sphere in a frame rotating at the microwave
frequency. When the gate charge $N_{g}$ varies as $N_{g}(t)=N_{g0}+\Delta
N_{g}\cos (2\pi \nu _{%
{\mu}%
w}t+\chi ),$ where $\chi $ is the phase of the pulse with respect to the
microwave reference, the hamiltonian $h=-\vec{H}.\overrightarrow{\sigma }/2$
is that of a spin 1/2 in an effective magnetic field $\vec{H}=h\Delta \nu \,%
\vec{z}+h\nu _{R0}\left[ \vec{x}\cos \chi +\vec{y}\sin \chi \right] $, where 
$\Delta \nu =\nu _{%
{\mu}%
w}-\nu _{01}$ is the detuning, and $\nu _{R0}=2E_{C}\Delta N_{g}\left\langle
1\right| \widehat{N}\left| 0\right\rangle /h$ the Rabi frequency. At $\Delta
\nu =0$, Rabi precession takes place around an axis lying in the equatorial
plane, at an angle $\chi $ with respect to the X axis.\ Rabi precession
between the qubit states induces oscillations of the switching probability $%
p $ with the pulse duration, as shown in Fig.~1, at a Rabi frequency that
scales with the amplitude~$\Delta N_{g}$~\cite{VionSCIENCE}. The range of
Rabi frequencies $\nu _{R0}$ that could be explored extends above $250\;%
\mathrm{MHz}$, and the shortest $\pi $ pulse duration for preparing $\left|
1\right\rangle $ starting from $\left| 0\right\rangle $ was less than 2 ns.
The fidelity was $\eta \simeq 0.3-0.4$ for readout pulses with $100~\mathrm{%
ns}$ duration.\ It was obtained with the readout performed at a value of the
phase $\langle \delta \rangle $ where the difference between the loop
currents for the two qubit states is maximised. The fidelity is smaller than
expected, possibly due to spurious relaxation of the qubit during the
readout process, and might be improved using rf methods that avoid switching
to the voltage state~\cite{Yale}.

In order to demonstrate that arbitrary operations on the qubit are possible,
it is necessary to combine rotations around different axes. In Fig~2,
measurements of the switching probability $p$ following\ two-pulse sequences
combining $\pi /2$ rotations around the axes $X,~Y,~-X$, or $-Y$, are shown.
The theory predicts that $p$\ oscillates at frequency $\Delta \nu $ with the
delay $\Delta t$\ between pulses. This experiment is analogous to the Ramsey
experiment in atomic physics, and to the free induction decay in NMR. When
the two pulses have different phases $\chi _{_{1}}$ and $\chi _{_{2}}$, the
Ramsey pattern is phase shifted by $\left( \chi _{_{2}}-\chi _{_{1}}\right) $%
.\ Despite the presence of spurious frequency jumps due to individual charge
fluctuators near the island, the overall agreement for the phase shift of
the Ramsey pattern demonstrates that rotations around axes $X$ and $Y$
combine as predicted. Arbitrary rotations of the qubit can thus be performed
by combining three rotations around these axes~\cite{NC}.\ However,
rotations around the $Z$\ axis can be more readily performed by changing the
qubit frequency for a short time.\ Note that frequency agility is also
interesting when interacting qubits need to be tuned at the same transition
frequency.\ In order to change the frequency, a triangular bias current
pulse with maximum amplitude $\Delta I$ is applied during a Ramsey
experiment.\ During the pulse, a phase difference $\zeta =2\pi \int \delta
\nu _{01}(t)dt$ builds up between the qubit states, which is equivalent to a
rotation around the $Z$ axis with an angle $\zeta $. The Ramsey pattern is
phase-shifted by $\zeta $ as shown by the oscillations of $p$ with $\Delta I$
(right panel in Fig~3).\ In the fit, the rotation angle is calculated using
the measured transition frequency as a function of $\delta $.\ Rotations
around the $Z$ axis have also been performed\ using adiabatic pulses in $%
N_{g}$.

Although these rotation experiments demonstrate that arbitrary unitary
transformations of the qubit state can be performed, their accuracy and
robustness are still issues. Similar questions have arisen in NMR, and
methods have been developed there to make spin manipulations less sensitive
to rf pulse imperfections. The transformation fidelity has furthermore been
addressed with the advent of NMR-based quantum computing~\cite{NC}. It has
been shown in particular that composite pulses can provide an appreciable
improvement~\cite{Jones}. In the case of the so-called $CORPSE$ sequence
(Compensation for Off-Resonance with a Pulse Sequence), a single pulse is
replaced by three, with the advantage of a strongly reduced sensitivity to
detuning.\ We have tested this sequence in the case of a $\pi $ rotation
around the $X$ axis, which performs a NOT operation on the qubit.\ The
corresponding $CORPSE$~$\pi (X)$ pulse sequence~is $\left\{ 7\pi /3(X),~5\pi
/3(-X),~\pi /3(X)\right\} $~\cite{Jones}. As shown in Fig.~4, this sequence
is significantly more robust against detuning than a single pulse $\pi (X)$
since the switching probability stays close to its maximum value over a
larger frequency span of about $100~\mathrm{MHz}$, comparable to the Rabi
frequency $\nu _{R0}=92~\mathrm{MHz}$.\ By performing the $CORPSE$ sequence
after an arbitrary rotation $\theta (-X),$ we have also checked that the $%
CORPSE$ sequence works for a general initial state.

The decay of Ramsey oscillations at long times provides a direct measurement
of the coherence time $T_{2}$.\ This decay was close to exponential with $%
T_{2}=300\pm 50~\mathrm{ns}$ for the present sample at the optimal working
point (see top panel in Fig.~5), and became progressively faster when
departing from that point~\cite{VionSCIENCE}.\ These observations validate
the concept of optimal working point, which has also been exploited recently
in flux qubits~\cite{DELFTprivCom}. The coherence time is limited by
relaxation and dephasing. Relaxation can be fought by better balancing the
two small josephson junctions~\cite{VionSCIENCE,ACthesis}. In the sample
reported in this work, the relaxation time $T_{1\text{ }}$was $500\pm 50~%
\mathrm{ns}$ at the optimal working point, where dephasing due to low
frequency charge and phase noises, which dominates decoherence, is minimum.
However, coherence times $T_{2\text{ }}$of the order of a few hundreds of
nanoseconds are still too short to implement quantum algorithms, and
increasing the coherence time is a major concern, for which concepts of NMR
are again useful. First, the well known spin-echo technique~\cite{nmr}
allows to suppress part of the dephasing. By inserting a $\pi $ pulse in the
middle of a Ramsey sequence, the random phases accumulated during the two
free evolution periods before and after the $\pi $ pulse cancel provided
that the perturbation is almost static on the time-scale of the sequence.\
The echo method provides a simplified form of error correction that
suppresses the effect of low frequency fluctuations.\ As shown in Fig.~5
(middle panel), the decay time of echoes is indeed longer than that of the
Ramsey pattern. At the optimal point, the echo decay time is roughly doubled
to $550\pm 50~\mathrm{ns}$.\ Away from the optimal point in the charge
direction $N_{g}$, this decay time is maintained at about $500~\mathrm{ns}$
till $T_{2}$ falls below $10~\mathrm{ns}$.\ Moving away in the phase
direction $\delta $, the echo compensation is less efficient, and the echo
decay time is roughly 2$T_{2}$ till $T_{2}$ falls below $20~\mathrm{ns.}$The
efficiency of echo compensation provides a probe of dephasing mechanisms,
and thus of noise sources.

Another way to increase the effective coherence time is to continuously
drive the qubit with a microwave signal, a method called spin-locking in NMR~%
\cite{nmr}. A spin-locking sequence consists of a Ramsey sequence, but with
a driving locking field along the direction $Y$ or $-Y$ continuously applied
between the two $\pi /2(X)$ pulses.\ Since the state $\left| -Y\right\rangle
=\left( \left| 0\right\rangle -i\left| 1\right\rangle \right) /\sqrt{2}$
prepared by the first pulse is then an eigenstate of the Hamiltonian in the
rotating frame in presence of the locking field, it does not evolve in time.
Relaxation between the states $\left| -Y\right\rangle $ and $\left|
Y\right\rangle $ occurs under the effect of fluctuations occuring at the
Rabi locking frequency, at a rate called the relaxation rate in the rotating
frame in NMR. \ As shown in Fig.~5 (bottom panel), the decay time of $p$
after the two spin-locking sequences $\left\{ \pi /2(X),~Lock(Y),~\pm \pi
/2(X)\right\} $, is equal to $650\pm 50~\mathrm{ns}$, which is significantly
longer than $T_{2}$.\ This decay time furthermore does not depend on the
orientation of the locking field along $Y$ or $-Y$\ because the energy
difference between the states $\left| Y\right\rangle $ and $\left|
-Y\right\rangle $ in the rotating frame is $h\nu _{R0}\ll kT$. Spin-locking
provides a weak form of error correction because low frequency fluctuations
of the hamiltonian are followed adiabatically by the eigenstates. This shows
that applying continuously a driving field can suppress part of decoherence
experienced by a qubit during its free evolution.\ 

In conclusion, we have demonstrated that the state of a quantronium qubit
can be efficiently \ manipulated using methods inspired from NMR. Rotations
around $X$ and $Y$ axes with microwave pulses have been combined, rotations
around the $Z$ axis have been performed with adiabatic pulses, and robust
rotations have been performed using composite pulses. Finally, the spin-echo
and spin-locking methods have yielded a significant increase of the
effective coherence time of the qubit. The quantitative investigation of
qubit decoherence using NMR techniques will be the subject of further work.

We acknowledge numerous discussions in the Quantronics group, the support of
the European projects SQUBIT 1 and 2, and of the program ''Action Concert%
\'{e}e Nanosciences''.


\begin{thebibliography}{99}
\bibitem{NC} M. A. Nielsen and I. L. Chuang, ``\textit{Quantum Computation
and Quantum Information}'', (Cambridge University Press, Cambridge, 2000).

\bibitem{Nak2} Y. Nakamura, Yu. A. Pashkin, J. S. Tsai, Nature \textbf{398},
786 (1999); Y. Nakamura \textit{et al.}, Phys. Rev. Lett. \textbf{88},
047901 (2002)

\bibitem{VionSCIENCE} D. Vion, A. Aassime, A. Cottet, P. Joyez, H. Pothier,
C. Urbina,\ D. Esteve, and M.H. Devoret, Science \textbf{296}, 886 (2002).

\bibitem{ACthesis} A.\ Cottet, PhD thesis, Universit\'{e} Paris VI, (2002);
www-drecam.cea.fr/drecam/spec/Pres/Quantro/ .

\bibitem{Martinis} J. M. Martinis \textit{et al.}, Phys. Rev. Lett. \textbf{%
89}, 117901 (2002).

\bibitem{delft} I. Chiorescu \textit{et al.}, Science \textbf{299}, 1869
(2003).

\bibitem{Chalmers} T. Duty \textit{et al.}, submitted to Phys. Rev. Lett.,
cond-mat/0305433.

\bibitem{Buisson} O.\ Buisson \textit{et al.}, submitted for publication.

\bibitem{NECgate} Yu. Pashkin \textit{et al.}, Nature \textbf{421}, 823
(2003).

\bibitem{quantroniumPHYSC} A. Cottet \textit{et al.,} Physica C \textbf{367}%
, 197 (2002).

\bibitem{Bouchiat98} V. Bouchiat \textit{et al.}, Phys. Scr. \textbf{T76},
165 (1998).

\bibitem{NakBOX} Y. Nakamura, C. D. Chen, and J. S. Tsai, Phys. Rev. Lett. 
\textbf{79}, 2328 (1997).

\bibitem{Yale} I. Siddiqi \textit{et al.}, submitted to Phys. Rev. Lett.,
cond-mat/0312623.

\bibitem{DELFTprivCom} Hans Mooij, private communication.

\bibitem{Jones} H.K.\ Cummins, G.\ Llewellyn, and J.A.\ Jones, Phys. Rev. A 
\textbf{67}, 042308 (2003).

\bibitem{nmr} C.P. Slichter, \textit{Principles of Magnetic Resonance},
Springer-Verlag (3rd ed: 1990).\newpage

\FRAME{ftbpFU}{10.0364cm}{7.3675cm}{0pt}{\Qcb{Top: circuit diagram of the
``quantronium''.\ The Hamiltonian of this circuit is controlled by the
gate-charge $N_{g}$ applied to the island between the two small junctions,
and by the phase $\protect\delta $\ across their series combination.\ This
phase is determined by the flux\ $\protect\phi $ through the loop, and by
the bias-current $I_{b}$.\ The two lowest energy states form a two level
system suitable for a quantum bit. The readout of the qubit state is
performed by inducing the switching of the larger readout junction to a
finite voltage V\ with a bias-current pulse $I_{b}(t)$ approaching the
critical current of this junction.\ Bottom: The quantum state is manipulated
by applying resonant microwave pulses (phase~$\protect\chi $) on the gate,
or adiabatic pulses on the bias-current. These pulses induce a rotation of
the effective spin $\protect\overrightarrow{\mathit{S}}$\ representing the
qubit state on the Bloch sphere in the rotating frame. The Rabi precession
of $\protect\overrightarrow{\mathit{S}}$ during a microwave pulse results in
oscillations of the switching probability $p$\ with the pulse length $%
\protect\tau $.}}{}{fig1.eps}{\special{language "Scientific Word";type
"GRAPHIC";maintain-aspect-ratio TRUE;display "USEDEF";valid_file "F";width
10.0364cm;height 7.3675cm;depth 0pt;original-width 5.9447in;original-height
5.8167in;cropleft "0";croptop "1";cropright "1.3367";cropbottom "0";filename
'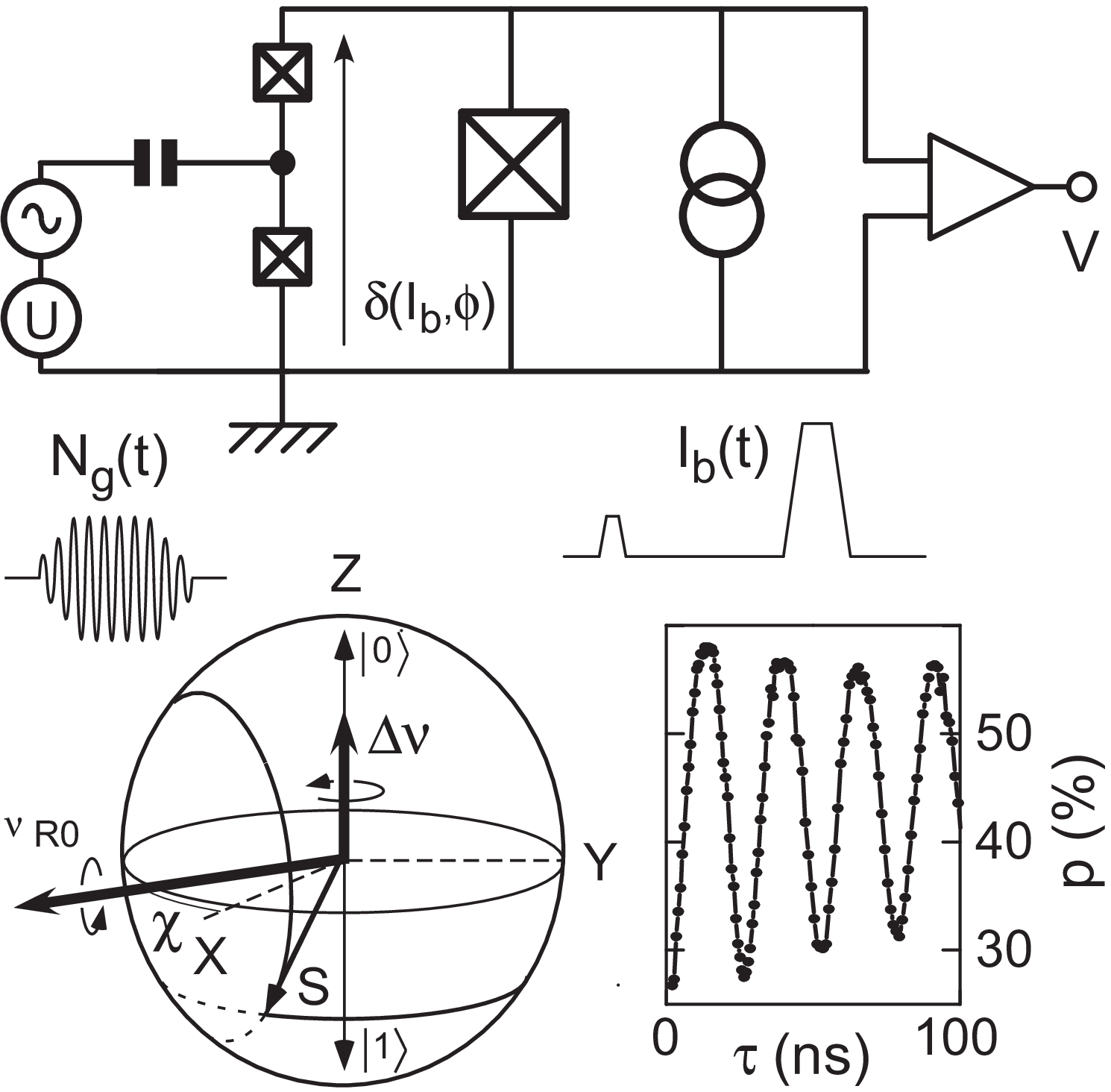';file-properties "XNPEU";}}\FRAME{ftbpFU}{10.0364cm}{%
9.7772cm}{0pt}{\Qcb{Switching probability after two $\protect\pi /2$\ pulses
with detuning $\Delta \protect\nu =+52~\mathrm{MHz}$, \ and with different
phases corresponding to rotation axes $X,Y,-X,$ or $-Y$, as a function of
the delay between the pulses.\ The solid lines are fits including a finite
decay time of 250 ns.\ The Ramsey patterns are phase-shifted as predicted
for the different combinations of rotation axes.}}{}{fig2c.eps}{\special%
{language "Scientific Word";type "GRAPHIC";maintain-aspect-ratio
TRUE;display "USEDEF";valid_file "F";width 10.0364cm;height 9.7772cm;depth
0pt;original-width 7.8646in;original-height 11.3334in;cropleft "0";croptop
"1";cropright "1.1087";cropbottom "0.2505";filename
'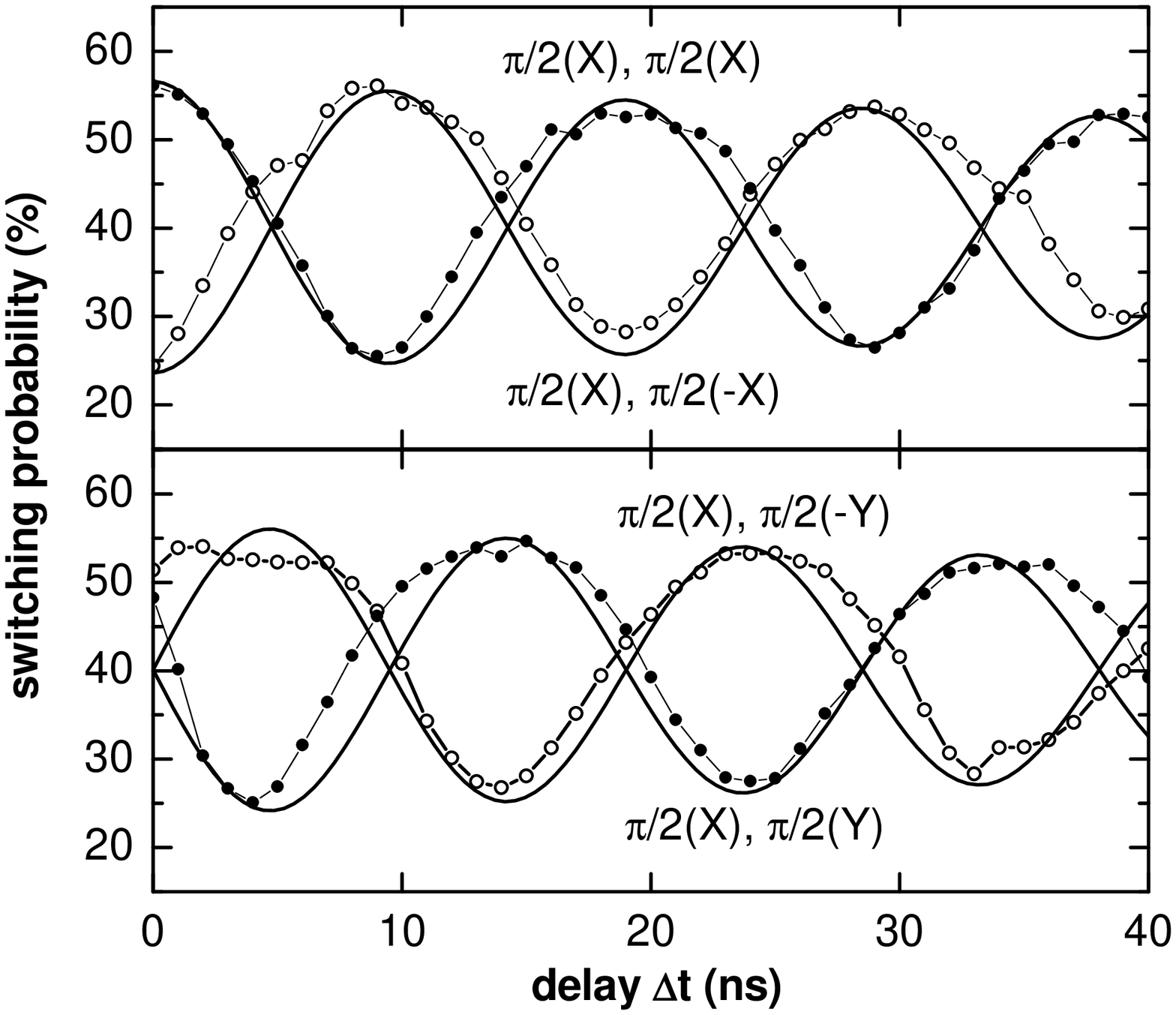';file-properties "XNPEU";}}\FRAME{ftbpFU}{11.0424cm}{%
6.658cm}{0pt}{\Qcb{Demonstration of rotations around the Z axis.\ A
triangular bias-current pulse applied between the two pulses of a Ramsey
sequence induces a frequency change, and thus a phase shift between the two
qubit states.\ This phase-shift, equivalent to a rotation around the $Z$
axis, results in oscillations of the switching probability $p$ (symbols)
with the pulse amplitude $\Delta I$.\ The fit uses the measured dependence
of the transition frequency with the phase $\protect\delta $.}}{}{fig3a.eps}{%
\special{language "Scientific Word";type "GRAPHIC";maintain-aspect-ratio
TRUE;display "USEDEF";valid_file "F";width 11.0424cm;height 6.658cm;depth
0pt;original-width 7.8646in;original-height 11.3334in;cropleft "0";croptop
"0.9122";cropright "1.1440";cropbottom "0.4354";filename
'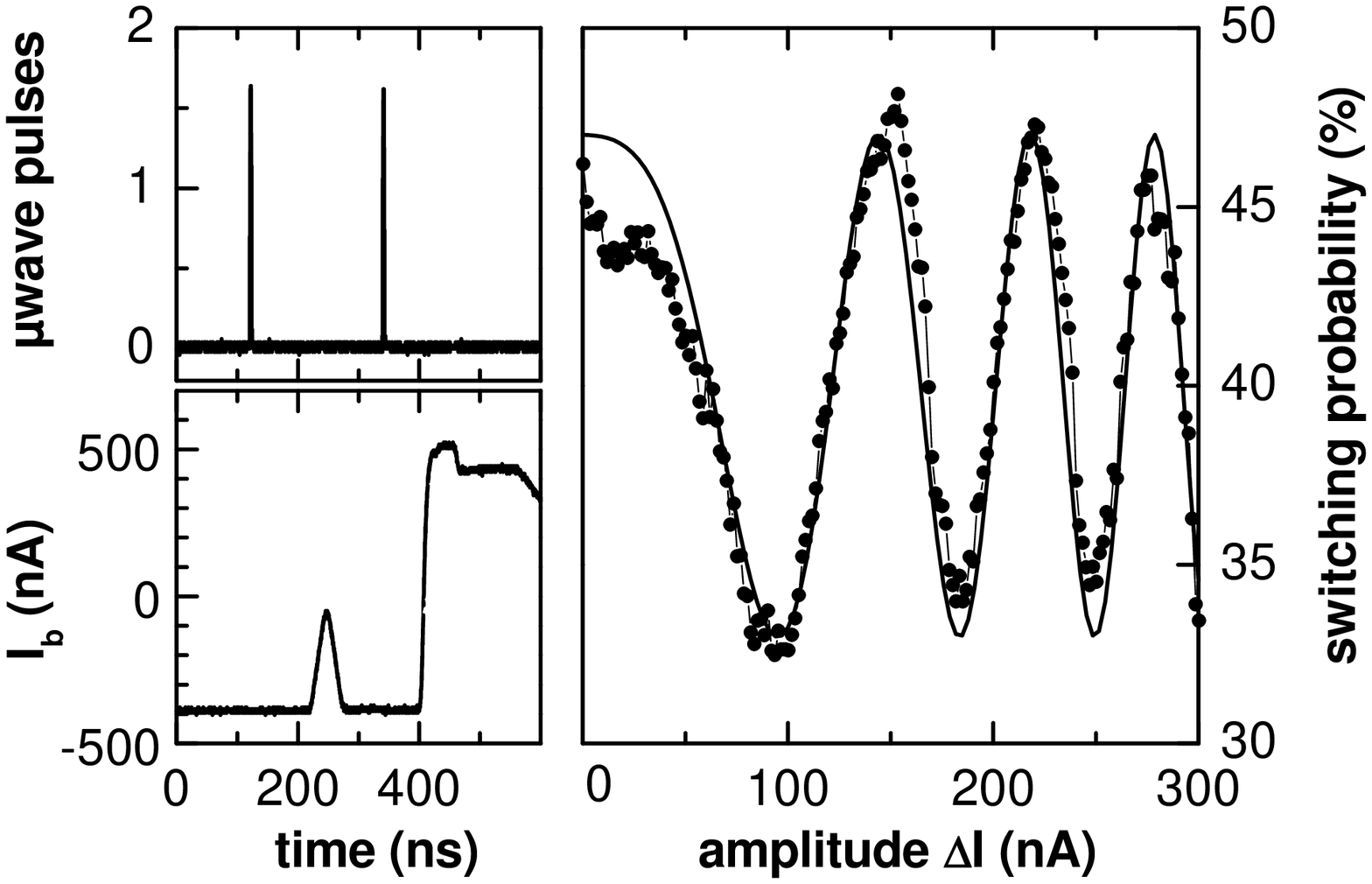';file-properties "XNPEU";}}\FRAME{ftbpFU}{11.5301cm}{%
6.5196cm}{0pt}{\Qcb{Demonstration of the robustness of a composite \ pulse
with respect to frequency detuning: Switching probability after a $CORPSE~%
\protect\pi (X)$ sequence(disks), and after a single $\protect\pi (X)$ pulse
(circles).\ The dashed line is the prediction for the $CORPSE~\protect\pi %
(X) $\ sequence, the arrow \ indicates the qubit transition frequency. The $%
CORPSE$ sequence works over a larger frequency range. The Rabi frequency was 
$92~\mathrm{MHz}$.\ Inset: oscillations of the switching probability after a
single pulse $\protect\theta (-X)$\ followed (disks) or not (circles) by a $%
CORPSE~\protect\pi (X)$\ pulse.\ The patterns are phase shifted by $\protect%
\pi $, which shows tthat the $CORPSE$ sequence does correspond to a not
operation.\ }}{}{fig4b.eps}{\special{language "Scientific Word";type
"GRAPHIC";maintain-aspect-ratio TRUE;display "USEDEF";valid_file "F";width
11.5301cm;height 6.5196cm;depth 0pt;original-width 7.8646in;original-height
11.3334in;cropleft "0";croptop "0.9124";cropright "1.2204";cropbottom
"0.4358";filename '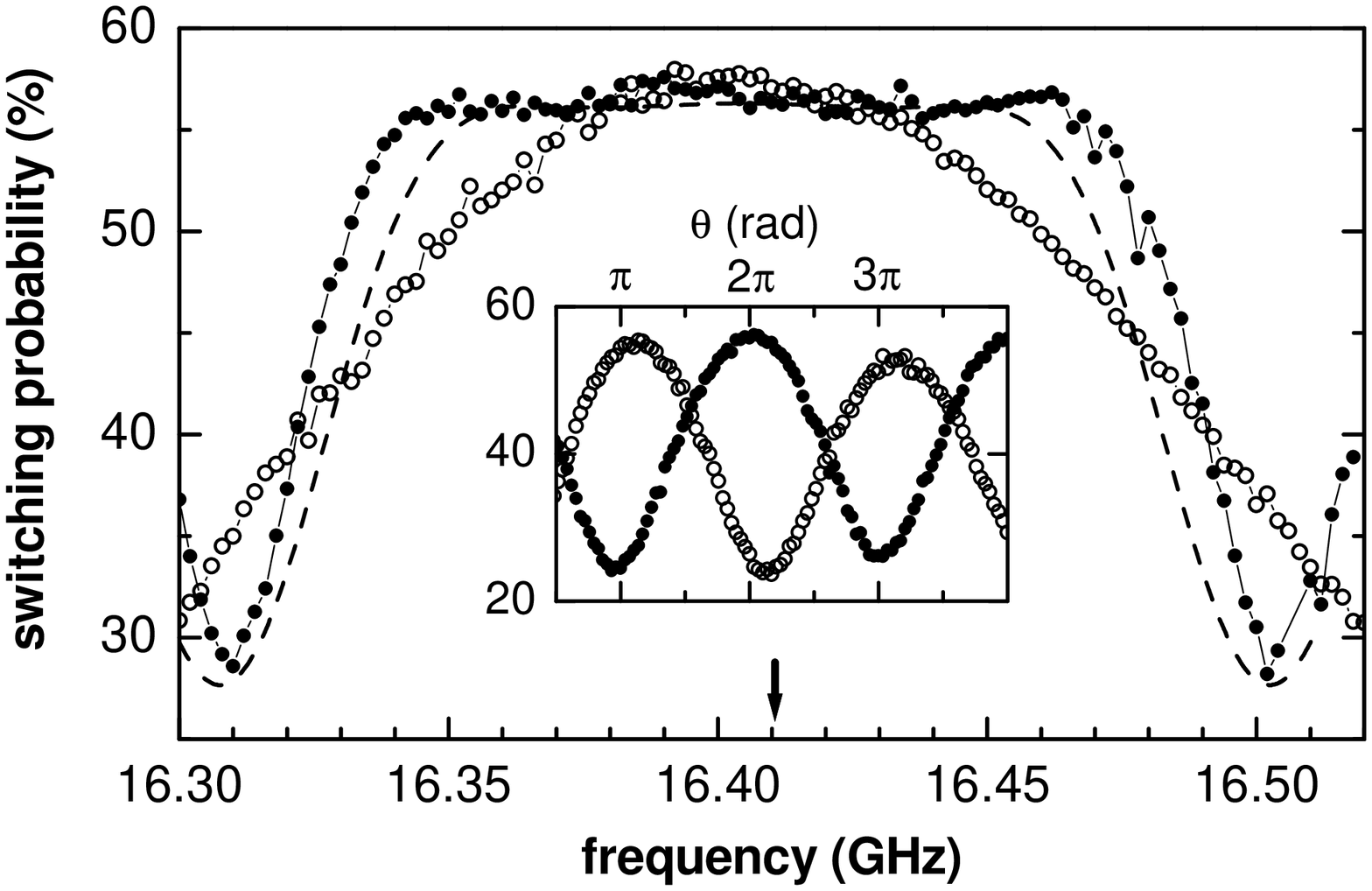';file-properties "XNPEU";}}\FRAME{%
ftbpFU}{10.4955cm}{12.4988cm}{0pt}{\Qcb{ Top panel: switching probability
(dots) after a Ramsey $\left\{ \protect\pi /2(X),~\protect\pi /2(X)\right\} $
sequence at $\Delta \protect\nu =+50~\mathrm{MHz}$, as a function of the
time delay between pulses. The lines are exponential fits of the envelope
with a time constant $T_{2}=350~\mathrm{ns}$. This decay time actually
varies due to changes in the charge fluctuators. Middle panel: example of
echo measured in a \ $\left\{ \protect\pi /2(X),~\protect\pi (X),~\protect%
\pi /2(X)\right\} $ sequence (dots).\ The arrow indicates the nominal
position of the echo minimum.\ Thin line: echo signal at the nominal minimum
position. The bold line is an exponential fit of the envelope with a $550~$ $%
\mathrm{ns}$ time constant. The dashed line shows a fit of the lower
envelope of the Ramsey pattern measured in the same conditions (220\ ns time
constant). Bottom panel: switching probability (thin lines) after two
spin-locking sequences with a Rabi locking frequency of $24~\mathrm{MHz,}$
at the optimal working point, as a function of the sequence duration. Thick
lines: exponential fits of the envelopes, with time constant $650~\mathrm{ns}
$. The dashed lines show a fit of the envelope of the Ramsey pattern
measured in the same conditions (time constant : 320\ ns). The residual
oscillations in echo and spin-locking signals are due to pulse sequence
imperfections.}}{}{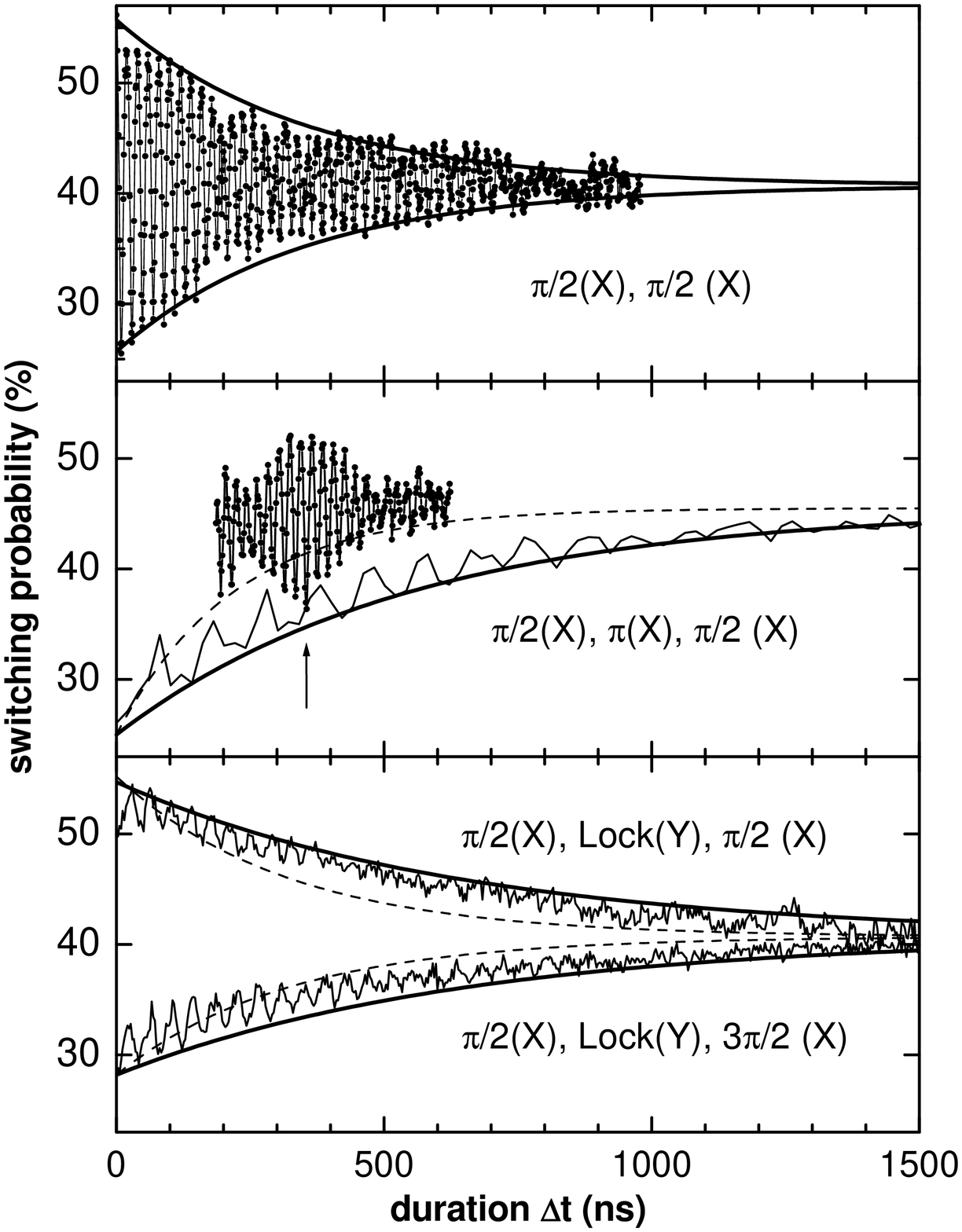}{\special{language "Scientific Word";type
"GRAPHIC";maintain-aspect-ratio TRUE;display "USEDEF";valid_file "F";width
10.4955cm;height 12.4988cm;depth 0pt;original-width 7.8646in;original-height
11.3334in;cropleft "0";croptop "1";cropright "1.1598";cropbottom
"0.0403";filename '../manuscrit/fig5e.eps';file-properties "XNPEU";}}
\end{thebibliography}
\end{document}